\documentclass[aps,prl,showpacs,floatfix,twocolumn]{revtex4}

\usepackage[dvips]{color}
\usepackage{mathptmx}
\usepackage[dvips]{epsfig}
\usepackage{amsmath}
\usepackage{amssymb}

\begin{document}

\title{Using high-power lasers for detection of elastic photon-photon scattering}  

\author{E.\ Lundstr\"om\footnote{Currently at: 
Department of Physics, Stockholm University, SE--106 91,
Sweden}, G.\ Brodin, J.\ Lundin, M.\ Marklund}
\affiliation{Department of Physics, Ume{\aa} University, SE--901 87 Ume{\aa},
Sweden}

\author{R.\ Bingham, J.\ Collier, J.T.\ Mendon\c{c}a, P.\ Norreys}
\affiliation{Rutherford Appleton Laboratory, Chilton, Didcot
OX11 0QX, Oxfordshire, UK}

\date{\today}

\begin{abstract}
  The properties of four-wave interaction via the nonlinear quantum vacuum 
  is investigated. The effect of the quantum vacuum is to generate photons with new
  frequencies and wave vectors, due to elastic photon--photon scattering. 
  An expression for the number of generated photons is derived
  and using state-of-the-art laser data it is found that the number of photons
  can reach detectable levels. In particular, the prospect of using the 
  high repetition Astra Gemini system at the Rutherford Appleton Laboratory
  is discussed. The problem of noise sources is reviewed, and it is found that 
  the noise level can be reduced well below the signal level. Thus, detection of elastic photon--photon scattering may for the first time be achieved.
\end{abstract}

\maketitle

Classically, elastic photon--photon scattering does not take place in vacuum. However, according to quantum electrodynamics (QED), such a process may occur owing to the interaction with virtual electron--positron pairs. Several suggestions to detect photon--photon scattering have been made, for example using harmonic generation in an inhomogeneous magnetic field \cite{Recent3}, using resonant interaction between eigenmodes in microwave cavities \cite{Brodin-Marklund-Stenflo,EBMS}, using ultra--intense fields occurring in plasma channels  \cite{Shen-Yu-Wang}, as well as many others, see e.g. Refs. \cite{Older2,Recent4,Recent6}. Related to, but physically different from, elastic photon--photon scattering is photon splitting \cite{Adler} (see also \cite{Gubbe}) and Delbr\"uck scattering \cite{Jarlskog}, the latter being detectable using high-$Z$ atomic targets \cite{Jarlskog}. 
However, no suggestions have yet led to detection of elastic scattering among real photons.

In the present Letter we will investigate the possibility to detect photon--photon scattering in vacuum using four--wave mixing, where three colliding laser pulses stimulate emission in a fourth direction with a new frequency. Four--wave mixing has the advantage of not being limited by the low cross section for photon--photon scattering \cite{Berestetskii}. A similar scheme was first studied by Ref. \cite{Adler70}, and further theoretical \cite{Recent,Moulin-Bernard,DiPiazza} and experimental \cite{Bernard-exp} studies of QED four--wave mixing have been performed since then. However, so far the available laser technology has not been sufficiently powerful to allow for successful detection of scattering events. Moreover, even with high laser power, the laser setup and geometry in such an experiment is important. In particular, we argue that a two--dimensional (2D) setup preferred in earlier literature is unlikely to produce a detectable signal. This is in contrast to the 3D setup presented in this Letter.  
We have calculated the coupling coefficient for four--wave mixing as a function of incident angles and laser polarization. For given data of the laser beams, together with the coupling coefficients, the number of scattered photons can be calculated. The analysis is then used to suggest a novel concrete experiment, where the parameters are chosen to fit the Astra Gemini (AG) system (operational 2007) located at the Central Laser Facility (CLF), Rutherford Appleton Laboratory (RAL). The estimated number of scattered photons per shot is 0.07. Furthermore, due to the polarization dependence of the number of scattered photons this experimental setup will allow for a unique fingerprint of the QED process. 
We note that a non--perfect vacuum in the interaction chamber leads to competing scattering events from Compton scattering and in principle to collective plasma four--wave mixing. Noting that plasma cavitation caused by the strong laser pulses will suppress the competing effects rather effectively close to the interaction region (see e.g. Ref. \cite{Guzdar}), we find the noise from competing scattering processes to be approximately three orders of magnitude lower than the QED signal in our suggested experimental setup. Thus, detection of photon--photon scattering will in principle be possible with the high--repetition rate AG system.

An effective field theory containing only the electromagnetic fields can be formulated in terms of the Heisenberg--Euler Lagrangian \cite{Heisenberg-Euler,Schwinger}, which is valid for field strengths below the QED critical field $10^{16}\,\mathrm{V\,cm^{-1}}$ and for wavelengths larger than the Compton wavelength $10^{-10}\,\mathrm{cm}$ \cite{general}. The Heisenberg--Euler Lagrangian is \cite{Berestetskii}
\begin{eqnarray}
  &&
  \!\!\!\!\!\!	{\cal L} 
   = ({8\pi})^{-1}\!\!\left\{ \!\!\left(E^{2}\!-B^{2}\right) + \xi\!\!\left[\left(E^{2}\!-B^{2}\right)^{2}+7\left(\textbf{E}\cdot\textbf{B}\right)^{2}\right]\!\! \right\}
\label{eq:lagrangian}
\end{eqnarray} 
where $\xi=\hbar e^{4}/45\pi m^{4}c^{7}$, $\hbar = h/2\pi$, $h$ is Planck's constant, 
$e$ is the magnitude of the electron charge, $m$ is the electron mass, and $c$
is the speed of light in vacuum. 
The first part is the classical Lagrangian density, while the second part represents the lower order non--linear QED--correction.  
From (\ref{eq:lagrangian}) follows 
\begin{equation}\label{eq:wave-equation}
	\Box\textbf{E} 
	= 4\pi \!\left[c^{2}\nabla \left(\nabla\cdot\textbf{P}\right) 
	- \partial_t\left(\partial_t\textbf{P} + c\nabla\times\textbf{M}\right)\right],
\end{equation}
where $\Box = \partial_t^{2} - c^2\nabla^{2}$. Here 
the effective polarization and magnetization are given by \cite{probing}
\begin{equation}\label{eq:polarization}
	\textbf{P}= (4\pi)^{-1}\xi\left[2\left(E^{2}\!-B^{2}\right)\textbf{E}+7\left(\textbf{E}\cdot\textbf{B}\right)\textbf{B}\right] 
\end{equation}
and
\begin{equation}\label{eq:magnetization}
\textbf{M}=  (4\pi)^{-1}\xi\left[-2\left(E^{2}\!-B^{2}\right)\textbf{B}+7\left(\textbf{E}\cdot\textbf{B}\right)\textbf{E}\right].
\end{equation}
Consider three plane waves, representing the incoming laser pulses, with amplitudes allowed to have a weak space time dependence, $\textbf{E}_{j}\!\left(\textbf{r},t\right) = (\tilde{\textbf{E}}_{j}\!\left(\textbf{r},t\right)e^{i\varphi_j(\mathbf{r},t)}+\text{c.c.})/2$, where c.c. denotes complex conjugate, $\varphi_j = \textbf{k}_{j}\cdot\textbf{r}-\omega_{j}t $ and $j=1,2,3$. 
Due to the cubic non--linearity of (\ref{eq:polarization}) and (\ref{eq:magnetization}), 
we expect generation of a fourth wave with $(\omega_{4},\textbf{k}_{4})$ if the incoming wave vectors satisfy the matching conditions
\begin{equation}\label{eq:wavevector}
	\textbf{k}_{1}+\textbf{k}_{2}=\textbf{k}_{3}+\textbf{k}_{4}\,\,\,\,\, \text{and}\,\,\,\,\,	\omega_{1}+\omega_{2}=\omega_{3}+\omega_{4},
\end{equation}
for some $\textbf{k}_{4}$ and $\omega_{4}\!=\!ck_{4}$.

We assume that the amplitudes change slowly, $\left|\partial E_{j}/\partial t\right| \ll \left|E_{j}\right|\omega_{j},\left|\nabla E_{j}\right| \ll \left|k_{j}\right|\left|E_{j}\right|$,
so the derivatives in the QED terms can be taken to act only on the harmonic parts. Only the resonant terms including the factors $\tilde{E}_{1}\tilde{E}_{2}\tilde{E}^{*}_{3}$ and $\tilde{E}^{*}_{1}\tilde{E}^{*}_{2}\tilde{E}_{3}$ are of importance, and hence the wave equation (\ref{eq:wave-equation}) for the generated field $\textbf{E}_{g}$ becomes
\begin{equation}\label{eq:wave-equation2}
	\Box{\textbf{E}}_{g}\!\left(\textbf{r},t\right)=4\xi\omega^{2}_{4}\textbf{G}\tilde{E}_{1}\tilde{E}_{2}\tilde{E}^{*}_{3}e^{i\left(\textbf{k}_{4}\cdot\textbf{r}-\omega_{4}t\right)}
\end{equation}
where 
$\textbf{G}$ is a geometric factor which only depends on the directions of the wave vectors and polarization vectors. The driving of the initial waves can be neglected due to the weakness of the generated field, and hence the  energy in each of these pulses is constant during the interaction.

The radiation zone solution to the wave equation (\ref{eq:wave-equation2}) is
\begin{equation}
{\textbf{E}}_{g}(\textbf{r},t)=\frac{\xi k^{2}_{4}}{\pi r}\textbf{G}e^{i\left({k}_{4}{r} -\omega_{4}t\right)}\int_{V'}\left.\left(\tilde{E}_{1}\tilde{E}_{2}\tilde{E}^{*}_{3}\right)\right|_{t_{R}}e^{i{k}_{4}\left(\hat{\textbf{k}}_{4}-\hat{\textbf{r}}\right)\cdot\textbf{r}'}dV',
\end{equation}
where $V'$ is the interaction region, and the amplitudes are to be evaluated at the retarded time $t_{R}=t-\left|\textbf{r}-\textbf{r}'\right|/c$.

The 2D and 3D geometric factors ($\textbf{G}_{2d}$ and $\textbf{G}_{3d}$ respectively) satisfying (\ref{eq:wavevector}), are given by long and complicated expressions \cite{Erik}. However, as will be demonstrated below, the polarization dependence of  $\textbf{G}_{3d}$ is important for the final design of our experiment. 

\begin{figure}[ht]
\includegraphics[width=0.8\columnwidth]{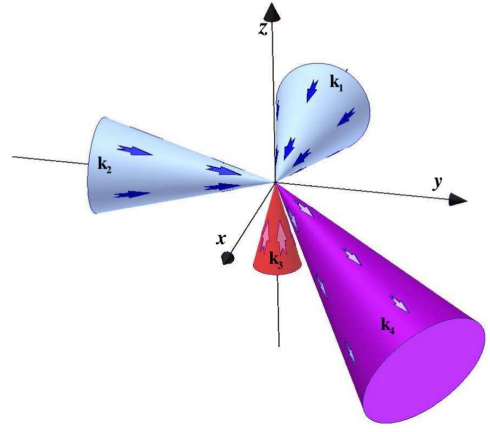}
\caption{Configuration of the incoming laser beams and the direction of the scattered wave for the suggested three-dimensional configuration of wave vectors, which satisfies the matching conditions (\ref{eq:wavevector}).}
\label{fig:3d}
\end{figure}

In practice, the highest power laser systems operates at a rather well--defined frequency. However, using frequency--doubling crystals we also have access to second harmonics, although with some power loss. As will be argued below, this makes a 2D geometry less competitive than a 3D setup. 
As a consequence, the 3D configuration of wave vectors which is best suited for an experimental setup (allowed by the matching conditions (\ref{eq:wavevector})) is given by $\textbf{k}_{1}=k\hat{\textbf{x}}$, $\textbf{k}_{2}=k\hat{\textbf{y}}$, $\textbf{k}_{3}= ({k}/{2})\hat{\textbf{z}}$, and $\textbf{k}_{4}=k\hat{\textbf{x}}+k\hat{\textbf{y}}-(k/2)\hat{\textbf{z}}$. This setup is illustrated in Fig.\ \ref{fig:3d}.

Here the wavelength and the direction of the generated wave is well separated from that of the incoming laser beams. This is highly desirable since optical components will always scatter and reflect some light, making it difficult to distinguish QED scattered photons of wavelengths close to that of the incoming laser beams. 
Furthermore, the geometry of the setup is not hindered by the fact that the focused laser beams are cone--shaped, and the polarizations can be chosen arbitrary since it is not necessary to prevent any counter propagating laser beam from entering back into the laser.

The number of scattered photons can be estimated from modeling the laser pulses as rectangular prisms with length $L$ and quadratic cross section $b^{2}$, inside which the field amplitudes are constant during the time of interaction. Although not a completely accurate model of the pulse shape, it will give a good estimate of the number of generated photons.

In the 3D case (see Fig.\ \ref{fig:3d}) the interaction region will take the shape of a cube with side $b$, existing during a time $L/c$. The generated electric field can now be solved for, from which the generated intensity and total power can be calculated. After multiplying the generated power by the interaction time and dividing by the photon energy $\hbar\omega_{4}$, the estimated number of scattered photons per shot is found to be
\begin{equation}\label{eq:N3d}
	N_{3d}=1.31\eta^{2} G_{3d}^{2}\left(\frac{1\,{\rm \mu m}}{\lambda_{4}}\right)^{3}\left(\frac{L}{1\,{\rm\mu m}}\right)\left(\frac{P_1P_2P_3}{1\,{\rm PW}^3}\right) ,
\end{equation}
where $G_{3d}$ is the geometric factor $|\mathbf{G}|$ evaluated for the given 3D geometry (see Ref.\ \cite{Erik}), $P_{j}$ is the power of the incoming pulses, $\lambda_{4}$ the generated wavelength, and with $\eta^{2}$ given by 
\begin{eqnarray}
  && \eta^{2} \equiv 
  \int^{2\pi}_{0}\int^{\pi}_{0}\frac{I_{\text{g}}\left(r,\theta,\phi\right)}{I_{\text{g, res}} 
  \left(r\right)}\sin\theta~d\theta ~d\phi
  \nonumber \\ && \!\!\!\!
  \approx 0.025\left(\frac{\lambda_{4}}{0.267\,{\rm \mu m}}\right)^2\left(\frac{1.6\,{\rm \mu m}}{b}\right)^2,
\end{eqnarray}
where $I_{\text{g}}$ is the generated intensity, and $I_{\text{g, res}}$ its maximum value in the resonant direction. The approximate equality of $\eta^{2}$ is accurate within $7\%$ for $b/\lambda_{4}<30$, i.e. $b/\lambda_{3}<10$ where $\lambda_{3}$ is the fundamental wavelength. 

\begin{figure}[ht]
\includegraphics[width=0.75\columnwidth]{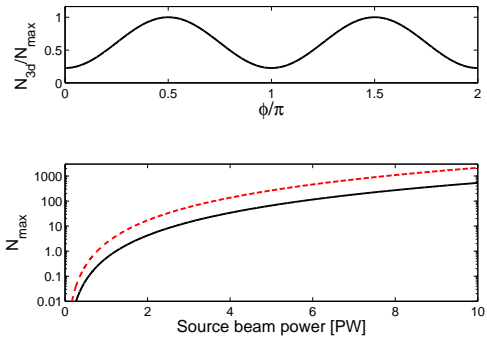}
\caption{The upper panel shows the number of scattered photons $N_{3d}$ per shot, normalized by the number of photons $N_{\text{max}}$ per shot for an optimal choice of polarization, as a function of polarization angle $\phi$ of the wave with direction $\hat{\mathbf{k}}_3$. 
The lower panel shows $N_{\text{max}}$  
predicted by (\ref{eq:N3d}) when increasing the laser power while keeping the beam width constant at $b=1.6\ \mathrm{\mu m}$, for a system where two source beams are used and one of the beams is split into two (solid line) and when three source beams are used, hence no beam splitting is required (dashed line). 
}
\label{fig:nr-of-photons}
\end{figure}

We note that the scaling of the number of generated photons with respect to $G_{3d}$ is decoupled from the pulse model. Hence it is possible to test the theoretical predictions by varying the polarizations of the incoming pulses, as seen in the upper panel of Fig.\ \ref{fig:nr-of-photons} where $N_{3d}$ is shown as a function of the polarization angle of wave three. 
For an optimal choice of polarization angles, $G^{2}_{3d} = 0.77$.

The AG Laser will generate two independently configurable 0.5 PW laser beams of wavelength $800$ nm. Each pulse will contain a total energy of 15 J, and focused intensities exceeding $10^{22}\ \mathrm{Wcm^{\!-2}}$ will be reached. The repetition rate is expected to be one shot per minute. Using these values, the spatial dimensions of the pulse model are taken as $b=1.6\ \mathrm{\mu m}$ and $L=10\ \mathrm{\mu m}$, which gives $\eta^{2}=0.025$. The 3D configuration described above can be achieved if one of the laser beams is frequency doubled and split into two beams. The estimated loss of power when frequency doubling is approximately 60\%, and hence the power of the incoming beams are given by $P_{1}=P_{2}=0.1$ PW, $P_{3}=0.5$ PW. The number of QED scattered photons, using optimal polarizations, would then be $N_{3d} =0.07$ and their wavelength around 267 nm. The lower panel of Fig.\ \ref{fig:nr-of-photons} shows how the scattering number increases with increasing laser power, when the focused beam width is kept constant. We note that in order to get a statistically sufficient number of scattered photons, the conditions must be reproduced from shot to shot. The pulse energy and length fluctuations will be small on a shot to shot basis and vibrations are likely to be a minor problem due to the inherent construction of the AG system. So, a good shot to shot reproducibility is expected from AG, and it can be further improved through use of comersially available active control systems. Below we will show that already the value $N_{3d}=0.07$ for the AG system will exceed the estimated noise level.

For comparison, let us consider a 2D configuration where two parallel beams collide head on with a third beam, all of them with the same wavelength. This configuration gives an optimal interaction region and also generates the most scattered photons, $N_{2d} =2.4$ per shot for an optimal choice of polarization angles with the AG system. However, the scattered photons are now emitted along the beam axis and have the same wavelength as that of the incoming beams, making detection impossible in practice. 

There are other possible experimental configurations that we have not investigated in this Letter. It is for instance possible to use configurations of two or more lasers of different wavelengths to generate QED scattered photons with wavelengths well separated from the harmonics of the incident beams, see Ref.\ \cite{Bernard-exp}. A 2D configuration consisting of a fundamental source beam frequency and its second harmonic will not give  
a scattering direction distinct from the incoming photon directions. Since present laser systems have sufficient power only at single frequencies (with second harmonics available using frequency--doubling crystals) we expect 2D configurations to be less competitive.

A perfect vacuum in the interaction chamber is in practice impossible to achieve. Below we give a rough estimate of the magnitude of the competing scattering events resulting from Compton scattering and collective plasma four--wave mixing, in order to determine whether it is possible to distinguish the QED scattered photons from noise.

We assume that we can filter out and detect only the photons within the wavelength interval $\pm 50$ nm centered at the wavelength of the QED scattered photons, $267$ nm. We calculate the scattering cross section for inverse Compton scattering into this wavelength interval in order to give a rough estimate of the noise caused by Compton scattered photons. For simplicity, the incident light is assumed to be monoenergetic with energy $\epsilon$. We further assume that in the rest frame of the electron, the scattering is elastic and isotropic Thomson scattering, which is a decent approximation as long as $\gamma\epsilon\ll mc^{2}$ \cite{lightman}, where $\gamma$ is the relativistic factor.

\begin{figure}[ht]
\includegraphics[width=0.75\columnwidth]{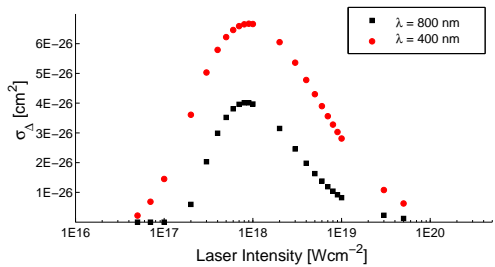}
\caption{The average cross section, $\sigma_{\Delta}$, for inverse Compton scattering of incident light with wavelength $800$ nm and $400$ nm into the wavelength interval $216-316$ nm, as a function of increasing intensity of the incident light.}
\label{fig:compton}
\end{figure}

Using the equations of motion,
$dp^{a}/d\tau=-(e/m)F^{a}\!_{b} p^{b}$,
the motion of an electron can be simulated and the average cross section for scattering into a given wavelength interval for a given field intensity can be estimated, see Fig.\ \ref{fig:compton}.
Here $p^a$ is the electron four-momentum, $\tau$ the electron eigen-time, and $F_{ab}$ is
the electromagnetic field tensor. 
Furthermore, below we investigate the effect of the ponderomotive force using a Gaussian laserpulse.

If the electrons were distributed homogeneously within a pulse of $N_{p}$ photons, the number of scattered photons during a time $\tau$ would be, $N_{\text{sc}}=\sigma n_{e} c N_{p} \tau$. However, of the scattered photons, only a fraction, $\delta$, will be scattered into the detector. Electrons accelerated by the laser pulse will most of the time have a relativistic velocity in the forward direction of the plane spanned by the polarization vector and the wave vector. From simulations, and taking into account the relativistic beaming effect, we estimate $\delta$ to be approximately $0.01$.
Of the three incoming pulses, only the $800$ nm beam will propagate away from the detector, see Fig.\ \ref{fig:3d}. From the above discussion and Fig.\ \ref{fig:compton}, we find that we can neglect the scattering from the $800$ nm beam. The number of photons in one beam is then about $N_{p}\approx6\times 10^{18}$ (40\% of 7.5 J at 400 nm).

With a maximum intensity of $10^{22}\ \mathrm{Wcm^{\!-2}}$ in the interaction region, we find from simulations that the effect of the ponderomotive force is crucial. We find that more than $99\%$ of the beam photons do not encounter any electrons, due to electron blow out, thus we introduce $\kappa=0.01$ reducing the scattering rate correspondingly. Of the scattered photons, only a fraction will have the energies possible to confuse with QED scattered photons (within that energy range we introduce the subscript $\Delta$ on the cross section $\sigma$). 
We find the average cross section $\sigma_{\Delta}=10^{-26}\ \mathrm{cm^{\!2}}$ from simulations (see Fig.\ \ref{fig:compton}). The expression for the number of Compton scattered photons per beam reaching the detector is then
\begin{eqnarray}
	N_{\text{det}}= \kappa \delta \sigma_{\Delta} n_{e} c N_{p} \tau. 
\end{eqnarray}

We assume the interaction region is imaged onto the detector in such a way that the detector only sees a part of the interaction region about $10\ \mathrm{\mu m}$ in diameter, which gives an interaction time of about $\tau\approx3\times10^{-14}\ \mathrm{s}$. We further assume an unperturbed vacuum around $10^{-9}$ torr, giving $n_{e}\approx 5\times10^{9}\ \mathrm{cm^{\!-3}}$. The number of Compton scattered photons that the detector might confuse with QED scattered photons, from both of the $400$ nm beams, is then about $N_{\text{det}}\approx5\times 10^{-5}$, about three orders of magnitude lower than the expected number of QED scattered photons.
Furthermore, because of the high vacuum and the electron blow out caused by the ponderomotive force, the collective plasma phenomena can be neglected compared to the Compton scattering.


In this Letter, we have investigated the possibility for QED elastic photon--photon scattering detection using four--wave mixing. The suggested experimental setup consisted of three colliding laser pulses in order to stimulate emission of a fourth electromagnetic wave with a new propagation direction and a new frequency. It has been shown that a 3D configuration, in contrast to previously suggested 2D setups, will be able to generate a measurable signal for the AG Laser and similar high power systems. Moreover, an application of the laser polarization makes it possible to obtain a fingerprint characterizing elastic photon--photon scattering.  
For given data of the laser beams, together with the coupling coefficients, an estimate of the number of scattered photons was obtained. The theoretical analysis was then used to suggest a concrete experiment, where the parameters were chosen to fit the next generation AG system at the CLF, RAL. The estimated number of scattered photons per shot was 0.07. Due to a non--perfect vacuum in the interaction chamber there is competing scattering events such as Compton scattering. However, due to plasma cavitation caused by the strong laser pulses the competing effects was shown to be suppressed rather effectively. Based on ultra-high vacuum technology it was found that the noise from competing scattering processes would be approximately three orders of magnitude lower than the QED signal in our suggested experimental setup. Thus, detection of photon--photon scattering will in principle be possible with the AG system operational in 2007.

\acknowledgments
E.L., G.B., J.L., and M.M.\ would like to thank The Centre for Fundamental Physics, RAL. 
The authors thank S.\ Hancock for help with illustration.
J.L.\ would further like to thank S. Bandyopadhyay, K. Ertel, P. Foster, and C. Murphy (CLF, RAL) for stimulating discussions.
This research was supported by the Swedish Research Council Contract
No.\ 621-2004-3217.



\begin{thebibliography}{99}

\bibitem{Recent3}  Y.J.\ Ding and A.E.\ Kaplan, J.\ Nonlinear Opt.\ Phys.\
Mater.\ \textbf{1} 51 (1992).

\bibitem{Brodin-Marklund-Stenflo}  G.\ Brodin, M.\ Marklund and L.\ Stenflo, Phys.\ Rev.\ Lett.\ \textbf{87} 171801 (2001).

\bibitem{EBMS}  D. Eriksson, G.\ Brodin, M.\ Marklund and L.\ Stenflo, Phys.\ Rev.\ A \textbf{70} 013808 (2004).

\bibitem{Shen-Yu-Wang}  B.\ Shen, M.Y.\ Yu and X.\ Wang, Phys.\ Plasmas \textbf{\ 10} 4570 (2003).

\bibitem{Older2}  E.B.\ Alexandrov, A.A.\ Anselm and A.N.\ Moskalev, Zh.\ Eksp.\ Teor.\ Fiz.\ \textbf{89} 1181 (1985) [engl.\ transl.\ Sov. Phys. JETP \textbf{62}, 680 (1985)].

\bibitem{Recent4}  M.\ Solja\v{c}i\'c and M.\ Segev, Phys.\ Rev.\ A \textbf{62}, 043817 (2000).

\bibitem{Recent6}  A.E.\ Kaplan and Y.J.\ Ding, Phys.\ Rev.\ A \textbf{62}, 043805 (2000).

\bibitem{Adler} S.L.\ Adler, Ann. Phys.-N.Y.\ \textbf{67}, 599 (1971).

\bibitem{Gubbe} J.D.\ Franson, Phys.\ Rev.\ A \textbf{53}, 3756 (1996); 
  \textit{ibid} \textbf{56}, 1800 (1997).

\bibitem{Jarlskog} G.\ Jarlskog et al., Phys.\ Rev.\ D  \textbf{8}, 3813 (1973).

\bibitem{Berestetskii} V.B.\ Berestetskii, E.M.\ Lifshitz and L.P.\ Ditaevskii, \textit{Quantum Electrodynamics} (Pergamon Press, Oxford, 1982).

\bibitem{Adler70}  S.L.\ Adler, J.N.\ Bahcall, C.G.\ Callan and M.N.\ Rosenbluth, Phys.\ Rev.\ Lett.\ \textbf{25}, 1061 (1970).

\bibitem{Recent}  N.N.\ Rozanov, Zh.\ Eksp.\ Teor.\ Fiz.\ \textbf{103} 1996 (1993) [engl.\ transl.\ Sov. Phys. JETP \textbf{76}, 991 (1993)].

\bibitem{Moulin-Bernard}  F.\ Moulin and D.\ Bernard, Opt.\ Comm.\ \textbf{164}, 137 (1999).

\bibitem{DiPiazza} A.\ Di Piazza, K.Z.\ Hatsagortsyan, and C.H.\ Keitel, Phys.\ Rev.\ D \textbf{72}, 085005 (2005).

\bibitem{Bernard-exp}  D.\ Bernard et al., Eur.\ Phys.\ J.\ D \textbf{10}, 141 (2000).

\bibitem{Guzdar}   G.Z.\ Sun, E.\ Ott, Y.C.\ Lee, and P.N.\ Guzdar, Phys.\ Fluids \textbf{30}, 526 (1987).

\bibitem{Heisenberg-Euler}  W.\ Heisenberg and H.\ Euler, Z.\ Physik \textbf{98} 714 (1936).

\bibitem{Schwinger}  J.\ Schwinger, Phys.\ Rev.\ \textbf{82} 664 (1951).

\bibitem{general}  Z.\ Bialynicka--Birula and I.\ Bialynicki--Birula, Phys.\ Rev.\ D \textbf{2} 2341 (1970).

\bibitem{probing} W.\ Dittrich and H.\ Gies, \textit{Probing the Quantum Vacuum} (Springer--Verlag, Berlin, 2000).

\bibitem{Erik} E. Lundstr\"om, M.Sc.\ Thesis (Ume\aa\ University, 2005), Appendix C, hep-ph/0512033.


\bibitem{lightman} G.B.$\!$ Rybicki and A.P.$\!$ Lightman, \textit{Radiation processes in astrophysics} (Wiley, New York, 1979).



\end{thebibliography}
\end{document}